\documentclass{jfm}
\usepackage{graphicx}
\usepackage{epstopdf}
\usepackage{epsfig}

\usepackage{subfigure}
\usepackage{amssymb}
\usepackage{amsmath}
\usepackage{mathrsfs}
\usepackage{placeins}
\usepackage{natbib}
\usepackage{caption3}
\bibpunct{(}{)}{,}{s}{,}{,}

\usepackage[bookmarksnumbered=true,breaklinks=true,colorlinks,citecolor=blue,linkcolor=blue]{hyperref}
\usepackage[geometry]{ifsym}
\makeatletter

\newcommand{\numtoRoman}[1]{\expandafter\@slowromancap\romannumeral #1@}

\renewcommand{\figurename}{Figure}

\makeatother

\shorttitle{Transition to turbulence in viscoelastic flow}
\shortauthor{A. Martinez Ibarra and J. S. Park}

\title{Transition to turbulence in viscoelastic channel flow of dilute polymer solutions}

\author{Alexia Martinez Ibarra
  \and Jae Sung Park
  \corresp{\email{jaesung.park@unl.edu}}}

\affiliation{Department of Mechanical and Materials Engineering, University of Nebraska-Lincoln,
Lincoln, NE 68588-0526, USA}

\begin{document}
\maketitle

\begin{abstract}
The transition to turbulence in a plane Poiseuille flow of dilute polymer solutions is studied by direct numerical simulations of a FENE-P fluid. A range of Reynolds number ($Re$) in $2000 \le Re \le 5000$ is studied but with the same level of elasticity in viscoelastic flows. The evolution of a finite-amplitude perturbation and its effects on transition dynamics are investigated. A viscoelastic flow begins transition at an earlier time than its Newtonian counterparts, but the transition time appears to be insensitive to polymer concentration at dilute or semi-dilute regimes studied. Increasing polymer concentration, however, decreases the maximum attainable energy growth during the transition process. The critical or minimum perturbation amplitude required to trigger transition is computed. Interestingly, both Newtonian and viscoelastic flows follow almost the same power-law scaling of $Re^\gamma$ with the critical exponent $\gamma \approx -1.25$, which is in close agreement with previous studies. However, a shift downward is observed for viscoelastic flow, suggesting that smaller perturbation amplitudes are required for the transition. A mechanism of the early transition is investigated by the evolution of wall-normal and spanwise velocity fluctuations and flow structure. The early growth of these fluctuations and formation of quasi-streamwise vortices around low-speed streaks are promoted by polymers, hence causing an early transition. These vortical structures are found to support the critical exponent $\gamma \approx -1.25$. Once the transition process is completed, polymers play a role in dampening the wall-normal and spanwise velocity fluctuations and vortices to attain a drag-reduced state in viscoelastic turbulent flows.
\end{abstract}


\section{Introduction}\label{sec:introduction}
Transition to turbulence has been studied extensively in wall-bounded shear flows for Newtonian fluids since the pioneering work of \cite{Reynolds1883a,Reynolds1883b}. However, despite many experimental and theoretical contributions \citep{Schmid2007,Eckhardt2007,Mullin2011,Avila2023}, its nature remains unclear even for simple geometries. \cite{Reynolds1883a,Reynolds1883b} noted in his experimental pipe flow studies that a strong perturbation can trigger transition at a Reynolds number ($Re$) of 2260. Subsequent studies have placed this critical Reynolds number in the range of $1760\leq Re\leq2300$ \citep{Kerswell2005}. Through more controlled conditions, it was also shown that the laminar state could be maintained until a higher Reynolds number of $Re=12,000$ (extended to $10^5$ by \cite{Pfenniger1961}). The upper critical $Re$ is in closer agreement with theoretical studies, showing that plane Couette flows (PCF) and pipe flows are linearly stable to infinitesimal perturbations for all Reynolds numbers \citep{DrazinReid1981}. For plane Poiseuille flows (PPF), experiments have shown a lower critical Reynolds number of $Re\approx1000$ \citep{DaviesWhite1928,OrszagKells1980,Carlson1982}, whereas linear stability theory has found that PPF becomes unstable at $Re\approx5772$ \citep{Orszag1971}. Experimental observations naturally point out the susceptibility of the flow to disturbances in the environment and explain why in practice most pipe and channel flows become turbulent at subcritical $Re$. In theory, this is further supported by analysis performed on the non-normality of the linearized Navier-Stokes equations, where the efficient amplification of finite-amplitude disturbances at a short time has been identified \citep{BobergBosa1988, Trefethen1993, SchmidHenningson2001}. In contrast to Newtonian flows, the onset of turbulence for non-Newtonian flows or viscoelastic flows of polymer solutions has been relatively less studied. 
In the remainder of this section, we provide a summary of the relevant literature concerning both Newtonian and viscoelastic flows and present the contributions addressed in the present study.

\subsection{Finite-amplitude thresholds in Newtonian flows}\label{subsec:newtonian}
Given the strong sensitivity of Newtonian flows to external disturbances, controlled perturbations have been widely utilized in the study of transitional flows. Of particular interest are the disturbances that cause the maximum energy growth in a specified time interval, known as linear optimal perturbations \citep{Farrell1988,ButlerFarrell1992}. For laminar pipe flow, the linear optimal disturbance is that of a counter-rotating streamwise vortex pair, which evolves into streamwise streaks due to the lift-up mechanism \citep{Landahl1980,SchmidHenningson1994}. These optimal perturbations are in agreement with the coherent structures that characterize transitional and turbulent shear flows. However, the transition is often triggered by other structures. \cite{Reddy1998} and \cite{PeixinhoMullin2007} showed that oblique disturbances are more successful at triggering turbulence. Hence, nonlinear optimization approaches have been proposed to compute optimal perturbations \citep{Monokrousos2011,Kerswell2018,LuchiniBottaro2014}, proving the existence and efficiency of nonlinear optimal perturbations over the linear ones \citep{PringleKerswell2010,Cherubini2013,Farano2015}. 

Of greater relevance to the current study is the study of the minimal perturbation amplitude $\epsilon$ required to trigger transition. The scaling law describing the relationship between $\epsilon$ and $Re$ is also of relevance. Early experimental work used continuous perturbations via a continuous injection of fluid through slits or holes \citep{WygnanskiChampagne1973, Rotta1956}. Impulsive perturbations, such as a single pulse injection of fluid, were also used, showing that these perturbations produced more consistent results by initiating controlled turbulent structures that could be used to determine turbulence far downstream \citep{Wygnanski1975, Rubin1980, DarbyshireMullin1995}. \cite{DarbyshireMullin1995} introduced various single-pulse disturbance configurations into a fully-developed pipe flow. The critical perturbation amplitude decreased very rapidly with increasing Reynolds number, eventually following an asymptotic behavior for high $Re$, regardless of the perturbation type. Hence, this behavior can be described as $\epsilon=O(Re^\gamma)$ with the critical exponent $\gamma < 0$, where a large $|\gamma|$ value corresponds to a rapid growth of the disturbances due to nonlinear effects \citep{Trefethen1993}.  The current estimate for the critical exponent $\gamma$ is in the range of $-7/4 \leq \gamma \leq -1$ derived from numerical and experimental studies for different geometries. For PPF, \cite{Lundbladh1994} and \cite{Reddy1998} performed numerical experiments and suggested $\gamma=-7/4$ for both streamwise and oblique perturbations. \cite{Chapman2002} used a formal asymptotic analysis of the Navier-Stokes equations and found $\gamma=-3/2$ and $\gamma=-5/4$ for oblique and streamwise initial perturbations. Experimentally, \cite{Philip2007} achieved an agreeable scaling factor of $\gamma=-3/2$ for PPF with a shorter channel length. For PCF, \cite{Dauchot1995} experimentally suggested the power exponent of $\gamma = -1$. Instead of using the perturbation amplitude, the kinetic energy of the perturbation $E_c$ has also been examined to suggest a similar scaling law for PCF, where $E_c = O(Re^{\gamma})$ with $-2 \le \gamma \le -2.7$ \citep{Kreiss1994,Duguet2013}. For pipe flow, \cite{Meseguer2003} numerically studied the formation and breakdown process of streaks due to streamwise vortices, suggesting a critical exponent of $\gamma=-3/2$, in agreement with the formal asymptotic analysis performed by \cite{Chapman2002} for PPF. Through novel experimental setups, \cite{Hof2003} and \cite{Lemoult2012} uncovered a scaling factor of $\gamma=-1$ for $Re>2000$, as proposed by \cite{WaleffeWang2005}. \cite{Lemoult2012} also showed an exponent close to $\gamma=-3/2$ for the restricted range of $1000 < Re < 2000$. Interestingly, \cite{MullinPeixinho2006} and \cite{PeixinhoMullin2006} showed that for $Re\leq1760$ turbulent flows cannot be sustained and all disturbances will eventually decay as $t \rightarrow \infty$.

\subsection{Transitional behaviour of drag-reducing flows}\label{subsec:viscoelastic}
Since the discovery of \cite{Toms1948}, the addition of small amounts of flexible long-chain polymers into a turbulent flow has been known to cause significant drag reduction (DR) in pipe and channel flows. This discovery attracted the interest of several applications that benefited directly from its drag-reducing effects. The most popular application of this phenomenon is in the fossil fuel industry (e.g. Alaska pipeline and fracking fluid). More recently, polymer additives were utilized in a large scale open-channel watercourse, which showed beneficial reduction in the water-depth downstream from the polymer injection point and an increase in the discharge capacity of the channel \citep{Bouchenafa2021}.

For viscoelastic effects, one of the most relevant non-dimensional numbers that characterizes polymer solutions is the Weissenberg number ($Wi$), which is the product of the longest relaxation time of the polymer solution and the characteristic shear rate of the flow. The other most relevant parameter is the elasticity number ($El = Wi/Re$), which is independent of the velocity, meaning that it is constant for a particular fluid and flow geometry. Hence, the DR phenomenon of polymer solutions in shear flows is typically described in terms of $Wi$ or $El$ \citep{Graham2014}.

The study of viscoelastic fluids has focused mainly on the drag-reducing phenomenon in a turbulent flow \citep{Min2003, WhiteMungal2008, Graham2014, Xi2019}, whereas the role of polymers on the onset of transition has been relatively less studied. Eariler pipe experments reported a lower transitional Reynolds number than one required for Newtonian transition, referred to as early turbulence \citep{RamTamir1964,Forame1972,Hansen1973,Zakin1977,Draad1998}. Recent experiments showed further possibilities of early transition in pipes and channels at sufficiently high polymer concentrations \citep{Samanta2013,SrinivasKumaran2017}, pointing at the influence of strong elastic effects on the onset of turbulence of polymer solutions. This turbulent state that results from early turbulence at high polymer concentrations is referred to as elasto-inertial turbulence (EIT) \citep{Samanta2013, Dubief2013,Terrapon2015,Sid2018}. \cite{Chandra2018} expanded the work of \cite{Samanta2013} for higher values of elasticity number and with various polymer types. For high polymer concentrations, they also found that transition occurred at $Re<2000$. This is in agreement with recent results of the linear stability theory of pipe flows by \cite{Garg2018} and \cite{Chaudhary2021}, who showed that pipe flows of an Oldroyd-B fluid are linearly unstable. However, it should be noted that \cite{Chandra2018} also observed the delayed transition, in other words, the transitional Reynolds number is increased. The study of EIT has also provided an alternative explanation to the upper limit of turbulent drag reduction, also known as the maximum drag reduction state \citep{Samanta2013, Choueiri2018, Lopez2019}. Interestingly, there are recent studies that have found the nonlinear elasto-inertial exact coherent structures in the EIT regime, named arrowhead structures \citep{Page2020, Dubief2022, Buza2022}, which link the EIT and elasto-inertial linear instability. An extensive review of these instabilities can be found on \cite{CastilloSanchez2022} and \cite{Datta2022}.

Similar to subcritical transition in Newtonian flows, a finite-amplitude perturbation is required to trigger the transition of polymer solutions. \cite{Hoda2008} studied the energy amplification of perturbations in the form of spatio-temporal body forces in PPF for an Oldroyd-B fluid. They found streamwise elongated disturbances to be the most amplified. \cite{Zhang2013} expanded this study to a finitely extensible nonlinear elastic fluid with the Peterlin closure (FENE-P) for inertia-dominated PPF. They observed the modal and non-modal types of perturbations, showing either stabilization or destabilization effects of polymer solutions depending on the polymer relaxation time. \cite{Agarwal2014,Agarwal2015} complemented these findings by spanning the bypass transition process for a FENE-P fluid in PPF. They observed the linear and nonlinear growth of an initially located disturbance and found a weakening of the disturbance amplification by polymers. A delay in the onset of transition and a prolonged transition period were also reported. For the natural or orderly transition of polymer solutions, \cite{LeeZaki2017} applied an infinitesimally small Tollmien-Schlichting wave to a FENE-P fluid in PPF. They found that the transition scenarios are affected by the level of the elasticity, where a destabilizing effect is observed at the lowest elasticity and a stabilization effect becomes manifested as the elasticity is further increased. \cite{Biancofiore2017} and \cite{Sun2021} investigated the nonlinear evolution of disturbed streaky structures in viscoelastic Couette and pipe flows, respectively, where viscoelasticity is found to delay the transition to turbulence in time for high $Wi$.

A power-law scaling of the critical perturbation amplitude, which is analogous to the Newtonian flow that relates $\epsilon$ and $Re$, has not been well explored for polymer solutions even at dilute concentrations and will be studied here. The transition of viscoelastic flows of a dilute FENP-P fluid is triggered by a finite-amplitude perturbation, and the effects of polymers on transition dynamics and mechanisms are reported. The problem formulation is reported in \S \ref{sec:formulation}. The simulation results are presented in \S \ref{sec:results}. We then conclude in \S \ref{sec:conclusion}.

\section{Problem Formulation}\label{sec:formulation}
We consider an incompressible fluid flow in the plane Poiseuille (channel) geometry driven by a constant mass flux. The $x$, $y$, and $z$ coordinates correspond to the streamwise, wall-normal, and spanwise directions, respectively. Periodic boundary conditions are imposed in the streamwise and spanwise directions with fundamental periods $L_x$ and $L_z$, respectively. No-slip boundary conditions are applied at the solid walls at $y = \pm h$, where $h$ is the half-channel height. Using the half-channel height $h$ and the Newtonian laminar centerline velocity $U_{cl}$ at the given mass flux as the characteristic length and velocity, respectively, the time $t$ is non-dimensionalized with $h/U_{cl}$ and pressure $p$ with $\rho U_{cl}^2$, where $\rho$ is the density of the fluid. Utilizing these characteristic scales, the nondimensional momentum and continuity equations for a fluid velocity $\boldsymbol{u}$ are
\begin{equation} \label{eq:ns}
\frac{\partial\boldsymbol{u}}{\partial t}+\boldsymbol{u}\cdot\nabla\boldsymbol{u}=-\nabla p+\frac{\beta}{Re}\mathrm{\nabla}^2\boldsymbol{u}+\frac{(1-\beta)}{Re}\nabla\cdot\boldsymbol{\tau}_p,\quad
\end{equation}
\begin{equation} \label{eq:cont}
\nabla \cdot \boldsymbol{u}=0
\end{equation}
Here, the Reynolds number for the given laminar centerline velocity is defined as $Re=\rho U_{cl}h/(\eta_s + \eta_p)$, where $(\eta_s+\eta_p)$ is the total zero-shear rate viscosity. The subscripts ‘$s$’ and ‘$p$’ represent the solvent and polymer contributions to the viscosity, respectively. The viscosity ratio $\beta=\eta_s/(\eta_s+\eta_p)$ (for a Newtonian fluid, $\beta = 1$). For dilute polymer solutions, $(1-\beta)$ is proportional to polymer concentration; hereinafter, the polymer concentration is represented as $c = 1 - \beta$. The concentration is assumed constant in time and homogeneous in space. Although the viscosity of polymer solutions displays shear-thinning, the total shear viscosity is hardly affected by the presence of the polymers for dilute solutions as polymers make a small contribution to the shear viscosity in the first place \citep{Graham2014}. The polymer stress tensor $\boldsymbol{\tau}_p$ is modelled by the FENE-P constitutive relation \citep{Bird1987} as 
\begin{equation} \label{eq:tau}
\boldsymbol{\tau}_p = \frac{1}{Wi} \left[ \frac{\boldsymbol{\alpha}}{1-\textrm{tr}(\boldsymbol{\alpha})/b}-\textbf{\textsf{I}} \right],
\end{equation}
where the Weissenberg number is defined as $Wi=\lambda U_{cl}/h$, where $\lambda$ is the polymer relaxation time. The parameter $b$ defines the maximum extensibility of the polymers (i.e., $\max{(\textrm{tr}(\boldsymbol{\alpha}))}\le b$), which is proportional to the number of monomer units. The polymer conformation tensor $\boldsymbol{\alpha}=\langle \boldsymbol{q}\boldsymbol{q} \rangle$ quantifies the second moment of the probability distribution for the polymer end-to-end vector $\boldsymbol{q}$, satisfying the evolution equation
\begin{equation} \label{eq:alpha}
\frac{\partial \boldsymbol{\alpha}}{\partial t}+ \boldsymbol{u}\cdot \nabla \boldsymbol{\alpha}-\boldsymbol{\alpha}\cdot \nabla \boldsymbol{u} -(\boldsymbol{\alpha}\cdot \nabla \boldsymbol{u})^{\textrm{T}} = -\boldsymbol{\tau}_p,
\end{equation}
which includes the upper convective derivative of $\boldsymbol{\alpha}$ and stress relaxation due to the elastic nature of the polymer.

Simulations are performed using the open-source code \textit{ChannelFlow} written and maintained by \cite{Gibson2012} from which a modified version was made and verified for viscoelastic flows used in the current study \citep{XiGraham2010, RoggePark2022}. This study focuses on results for the range of $2000\le Re\le5000$. This Reynolds number range for Newtonian flows is found to be subcritical and below the linear stability limit for two-dimensional flows but slightly beyond the transition for three-dimensional flows \citep{SchmidHenningson2001}. For the viscoelastic cases, the polymer concentration ranges from dilute to semi-dilute regimes: $0.01\le c\le0.09$. The Weissenberg number ranges from $32\lesssim Wi\lesssim65$. Note that the current study holds the elasticity number constant at $El \approx 0.017$. The parameter $b = 5,000$, which corresponds to a moderately flexible, high-molecular-weight polymers \citep{XiGraham2010, XiGraham2012}. The extensibility parameter $Ex$ is defined as the polymer contribution to the steady-state stress in uniaxial extensional flow. For the FENE-P model, $Ex=2b(1-\beta)/3\beta$. Significant effects of polymers on turbulence are only expected when $Ex \gg 1$ for a dilute solution $(1-\beta \ll 1)$ which is the case of this study. For the sets of $\beta$ and $b$ studied, the values of $Ex$ are in the range of $34 - 330$, which is sufficient to observe the effects of polymer solutions \citep{XiGraham2010}. This parameter space for the viscoelastic flow is found to be linearly stable \citep{Datta2022, CastilloSanchez2022}.

The equation system above is coupled and integrated in time with a third-order semi-implicit backward differentiation and Adams-Bashforth method for the linear and nonlinear terms, respectively \citep{Peyret2002}. As an effective approach to identifying the self-sustain process in both Newtonian and viscoelastic flows \citep{JimenezMoin1991, Webber1997}, the so-called minimal flow unit (MFU) is employed. We use a domain of $L_x\times L_y\times L_z=2\pi\times2\times\pi$ and $4\pi\times2\times2\pi$ to simulate Newtonian and viscoelastic flows, respectively. It is worth noting that viscoelastic MFUs are larger than Newtonian ones to attain sustained turbulence \citep{WangGraham2014}. A numerical grid system is generated on $(N_x,N_y,N_z)$ (in $x$, $y$, and $z$) meshes, where a Fourier-Chebyshev-Fourier spectral spatial discretization is applied to all variables and nonlinear terms are calculated with the collocation method, for which the standard $2/3'$s dealiasing is used. The numerical grid systems used are $(N_x,N_y,N_z) = (64,81,76)$ for the Newtonian simulations and $(N_x,N_y,N_z) = (126,81,126)$ for the viscoelastic simulations, unless specified otherwise. The numerical grid spacings in the streamwise and spanwise direction are uniform with $\Delta x^+\approx 12$ and $\Delta z^+\approx 7$, respectively, for all cases. In the wall-normal direction, the non-uniform Chebyshev spacing is $\Delta y_{min}^+\lesssim0.1$ at the wall and $\Delta y_{max}^+\approx 5$ at the channel centre.

An artificial diffusivity term $1/(ScRe)\nabla^2\boldsymbol{\alpha}$ with the Schmidt number $Sc = 0.5$ is added to the right-hand side of eq.~\ref{eq:alpha} to improve its numerical stability, as is common practice for spectral simulations of viscoelastic flows \citep{Sureshkumar1997, Dimitropoulos1998, XiGraham2010, XiGraham2012, RoggePark2022}. For the range of $Re$ given above, an artificial diffusivity $1/(ScRe)$ is of the order of $10^{-3}-10^{-4}$, which is much lower than often used in other studies with $1/(ScRe) = O(10^{-2})$ \citep{Sureshkumar1997, Ptasinski2003, Li2006, Li2015}. In the low-to-moderate $Wi$ and dilute-to-semi-dilute regimes of the present study, these very small magnitudes of artificial diffusivity should not have a significant impact on the numerical solutions, while still contributing to numerical stability, which have also been confirmed by previous studies \citep{Sureshkumar1997, Housiadas2005, Li2006, Kim2007, ZhuXi2020}. Since introducing such an artificial term, an additional treatment for a boundary condition on eq.~\ref{eq:alpha} is needed. We update $\boldsymbol{\alpha}$ at the walls using the solution without the artificial diffusivity. These results are then used as the boundary condition to solve eq.~\ref{eq:alpha} with the artificial diffusivity term and update $\boldsymbol{\alpha}$ for the rest of the channel. The numerical details used in the present study can be found in \cite{Xi2009}. The numerical code used here has been extensively validated in the previous studies \citep{XiGraham2010, XiGraham2012, WangGraham2014, WangGraham2017, RoggePark2022}.

The initial velocity field is a superposition of the parabolic laminar base flow $\boldsymbol{u}_{lam}$ and a three-dimensional perturbation flow: $\boldsymbol{u} \boldsymbol = \boldsymbol{u}_{lam} + a\boldsymbol{u}_{p}$, where $a$ is the magnitude of the perturbation flow field, which is adjustable in the current study. Different laminar base flow $\boldsymbol{u}_{lam}$ is used for Newtonian and viscoelastic flows. The Newtonian laminar flow is a typical parabolic velocity profile of a plane Poiseuille flow \citep{White2006}, while a viscoelastic laminar flow is obtained from the plane Poiseuille flow solution of a FENE-P fluid \citep{Cruz2005}. The viscoelastic laminar flow is a modified version of the Newtonian laminar flow to which contributions due to polymers (i.e., $Wi$, $\beta$, and $b$) are added. In addition, the laminar base state for the polymer stress tensor is also considered \citep{LeeZaki2017}. The perturbation field $\boldsymbol{u}_{p}$ is generated using the subroutine \textit{randomfield} through \textit{ChannelFlow} \citep{Gibson2012}, where its spectral coefficients of three velocity components are set to decay exponentially with respect to the wavenumber to ensure the smoothness of the flow similar to turbulent fields. The perturbation field also satisfies no-slip and divergence-free conditions (see appendix C in \cite{Pershin2022} for details of the subroutine \textit{randomfield}). This perturbation field is similar to that commonly used for the optimal disturbance to control a transition to turbulence \citep{Farano2015, Pershin2022}. However, it should be emphasized that the particular form of $\boldsymbol{u}_p$ does not matter as long as it leads to an instability that triggers a transition to turbulence \citep{FaisstEckhardt2004}. Owing to the extensional flow nature of transitional and turbulent flows, there are always positive Lyapunov exponents in Newtonian channel flows \citep{Keefe1992, Nikitin2018} and even viscoelastic channel flows \citep{StoneGraham2003}, resulting in a quick memory loss of the initial conditions. \cite{DarbyshireMullin1995} also experimentally confirmed that different kinds of perturbations result in a very similar stability curve. Nonetheless, the choices of the different forms of the perturbation field were tested, showing similar behaviours such as scaling laws \citep{Mullin2011}. In addition, for optimal perturbations, where the maximum energy growth is efficiently reached during the transition process, similar scaling behaviour was observed \citep{Farano2015}. Therefore, it can be safely assumed that the effect of the perturbation field on the transition to turbulence can be focused only on its magnitude.

Throughout the paper, the perturbation amplitude $A$ is defined as the ratio of the $L_2$-norm of the perturbation velocity field $\boldsymbol{u}_{p}$ to that of the base laminar velocity field $\boldsymbol{u}_{lam}$:
\begin{equation}
A = \frac{|| \boldsymbol{u}_{p}||_2}{|| \boldsymbol{u}_{lam}||_2} = a \sqrt{ \int_{V}\boldsymbol{u}_{p}^2~dV \bigg/ \int_{V}\boldsymbol{u}_{lam}^2~dV },
\end{equation}
where $V = 2L_x L_z$ is the volume of the computation domain. The amplitude square $A^2$ can also be referred to the ratio of the kinetic energy of $\boldsymbol{u}_{p}$ to that of $\boldsymbol{u}_{lam}$. The perturbation amplitude studied is in the range of $0.014 \leq A \leq 0.14$ to ensure small-amplitude perturbations for promoting a linear instability during the transtition to turbulence \citep{SchmidHenningson2001}. It is also important to note that due to the addition of a global artificial diffusion and use of spectral method, it is almost unachievable to trigger transition to turbulence for $Re<2000$ even with a sufficiently large perturbation amplitude $A > 0.14$.

Prior to proceeding to the results, it is worth emphasizing the flow regime of the current study. For $El < 0.02$ and $2000 \leq Re \leq 5000$, the flow regime of interest can be referred to as inertia-driven transition for both Newtonian and dilute viscoelastic flows \citep{Datta2022}. The EIT flow regime, which is typically $El\gg 0.02$ \citep{Samanta2013, Dubief2013}, is distinctly different from the current flow regime. Thus, a quantitative or even qualitative comparison between the current inertia-driven transition and EIT transition should not necessarily be expected in the following results.

\section{Results and discussion}\label{sec:results}
\begin{figure}
\begin{center}\includegraphics[width = 5in]{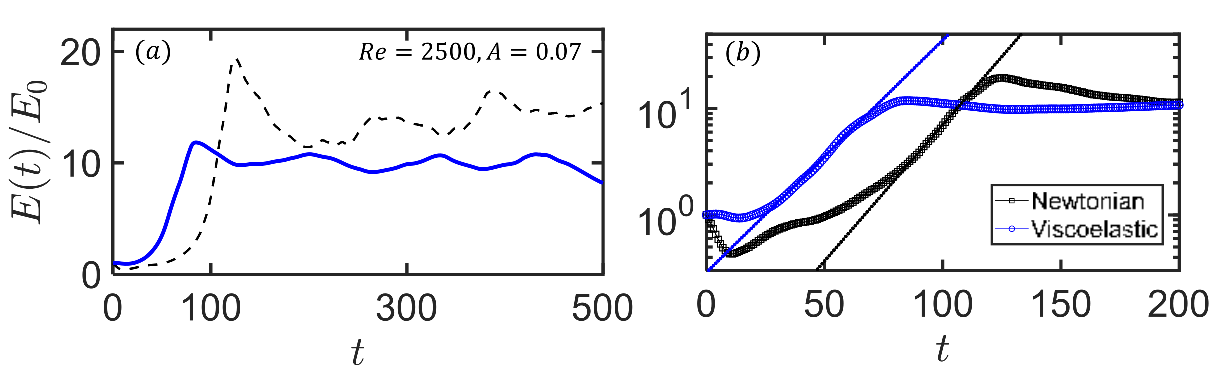}
\caption{Temporal evolution of the disturbance energy $E(t)$ normalized by the initial disturbance energy $E_0$ on ($a$) a linear-linear scale and ($b$) a log-linear scale for Newtonian (black dashed line/squares) and viscoelastic flow ($c = 0.03$ or $\beta = 0.97$; blue solid line/circles). Both flows are disturbed by the perturbation amplitude $A = 0.07$ at $Re = 2500$. The amplification of the perturbation (straight lines in $b$) behaves like $e^{\sigma t}$, where $\sigma = 0.06$ and $0.05$ for Newtoninan and viscoelastic flows, respectively. 
\label{fig:energyvn_gr}}
\end{center}
\end{figure}

We perform direct numerical simulations starting from a laminar base flow disturbed with a small finite-amplitude perturbation for both Newtonian and viscoelastic flows. The amplitude of the perturbation was set to be in the range of $0.014\leq A \leq0.14$ relative to the total energy of the laminar base flow.

\begin{figure}
\begin{center}\includegraphics[width = 5in]{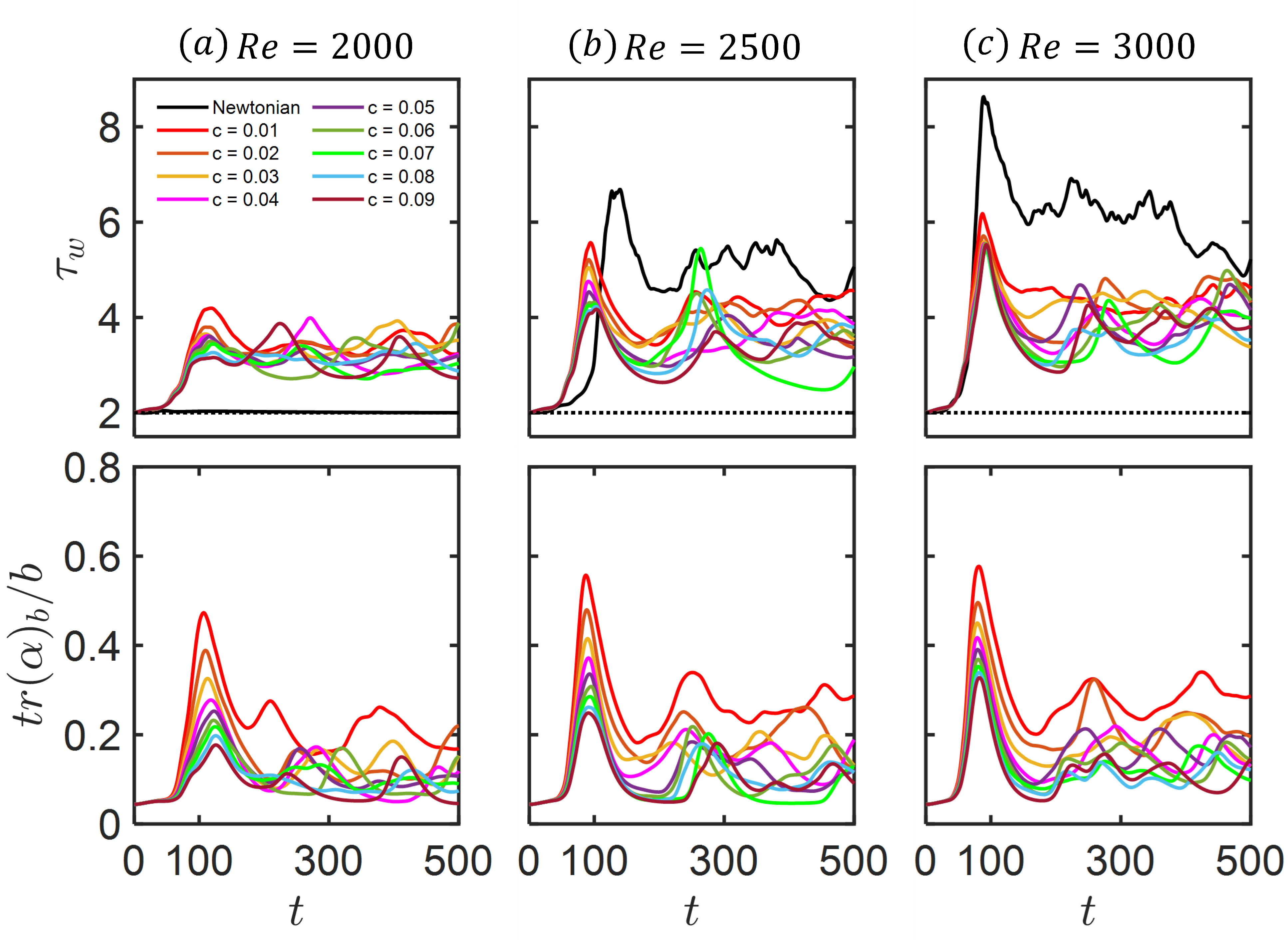}
\caption{Temporal evolution of the (top panel) wall shear stress $\tau_w$ and (bottom panel) bulk polymer stretching normalized by the maximum extensibility of polymers $tr(\alpha)_b/b$ for the perturbation amplitude $A=0.07$ at (a) $Re=2000$ (b) $Re=2500$, and (c) $Re=3000$: laminar state (black dashed line), Newtonian flow (black solid line), and viscoelastic flows of various polymer concentrations (coloured solid lines).
\label{fig:wallshearstress}}
\end{center}
\end{figure}

\subsection{Transition dynamics}\label{subsec:transition}
To examine the temporal behaviour of the transition dynamics, figure~\ref{fig:energyvn_gr}($a$) illustrates the evolution of the disturbance energy per unit volume $E(t)$, which is given as
\begin{equation}
E(t)=\frac{1}{2L_xL_z}\int_{0}^{L_z}\int_{-1}^{1}\int_{0}^{L_x} (u'^2+v'^2+w'^2)\textrm{d}x\textrm{d}y\textrm{d}z,
\end{equation}
where the streamwise velocity fluctuation $u' = u - u_{\textrm{lam}}$, while $v' = v$ and $w' = w$ since $v_{\textrm{lam}} = 0$ and $ w_{\textrm{lam}} = 0$. Profiles are normalized by the initial disturbance energy $E_0$. At $t = 0$, the same amplitude of perturbation $A = 0.07$ was applied to both Newtonian and viscoelastic ($c=0.03$ or $\beta=0.97$) flows at $Re = 2500$. As has been typically observed in transition to turbulence \citep{SchmidHenningson2001}, both flows exhibit a similar early-time behaviour of the energy growth: (i) an initial stable period, (ii) a sharp increase up to the maximum value, or namely a strong burst, and (iii) transition to a fully-turbulent flow. However, the first notable distinction between both flows is the duration of the initial stable period. As clearly seen in figure~\ref{fig:energyvn_gr}($a$), the viscoelastic flow experiences a shorter stable duration than the Newtonian counterpart. In other words, polymers appear to destabilize the flow earlier than the Newtonian flow, triggering an earlier transition. Another distinction lies in the strong burst whose magnitude is significantly reduced by polymers. This strong burst has also been referred to as the escaping process out of the so-called exact coherent solution along its most unstable manifold \citep{Itano2001,Park2018}, comprising of the linearly unstable stage followed by the nonlinear evolution stage. A log-linear representation of the evolution of the disturbance energy per unit volume $E(t)$ is shown in figure~\ref{fig:energyvn_gr}($b$). The exponential amplification of the perturbations in both Newtonian and viscoelastic flows is marked by dotted lines with $\sigma$ values equal to $0.06$ and $0.05$ for Newtonian and viscoelastic flows, respectively. After the strong burst, both flows enter a fully-turbulent regime at $t \approx 150$, where turbulent drag reduction via polymers is manifested.

\begin{figure}
\begin{center}\includegraphics[width = 5.4in]{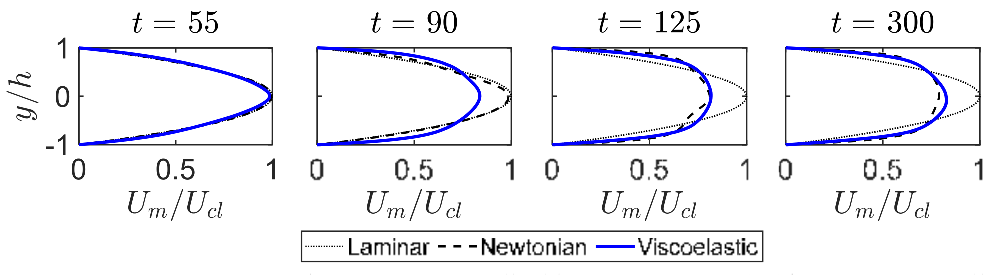}
\caption{Temporal evolution of the mean velocity profile $U_m$ normalized by the laminar centerline velocity $U_{cl}$ for the perturbation amplitude $A=0.07$ at $Re=2500$ with the time instants $t = 55, 90, 125,$ and $300$: laminar state (black dotted line), Newtonian flow (black dashed line), and viscoelastic flow ($c=0.03$; blue solid line).
\label{fig:velprofs2}}
\end{center}
\end{figure}

To further characterize the transition to turbulence, we utilize the mean wall shear stress $\tau_w$ as means to estimate the transitional trajectory of Newtonian and viscoelastic flows, as it has been widely utilized to characterize the intermittent dynamics of both flows \citep{XiGraham2012}. The top panel in figure~\ref{fig:wallshearstress} shows the temporal evolution of the wall shear stress for Newtonian and viscoelastic flows of various polymer concentrations perturbed with $A = 0.07$ at $Re = 2000$, $2500$, and $3000$ along with the base laminar state whose wall shear stress $\tau_{w,lam} = 2$. The response of the flow to the perturbation can be divided in two distinct cases: (a) the flow remains undisturbed or (b) the flow departs from the base laminar state and begins its path to turbulence shortly after the introduction of the perturbation. For $Re = 2000$ (fig.~\ref{fig:wallshearstress}$a$), the Newtonian flow remains undisturbed (case a), whereas the viscoelastic flows of all polymer concentrations begin transition at a few time instants after the introduction of the perturbation (case b). For $Re\geq2500$, both Newtonian and viscoelastic flows follow a transition trajectory (case b). As clearly seen in figure~\ref{fig:wallshearstress}($b$), the transition begins at an earlier time for viscoelastic flows than for their Newtonian counterpart, indicating an earlier transition. At a higher Reynolds number of $Re = 3000$ (fig.~\ref{fig:wallshearstress}$c$), however, the transition appears to begin at almost the same time for Newtonian and viscoelastic flows. Similar to the perturbation energy in figure~\ref{fig:energyvn_gr}($a$), the maximum wall shear stress of the strong burst in viscoelastic flows is smaller than the one in Newtonian flow. Furthermore, the maximum wall shear stress is reduced with increasing polymer concentration, implying that the magnitude of the strong burst decreases with polymer concentration. The reduction of the aforementioned maximum wall shear stress of the polymer solutions is also an important indicator of drag reduction in sustained turbulent flow regimes. Hence, turbulent drag reduction is expected for viscoelastic flows compared to Newtonian counterparts, as can be seen in the wall shear stress in figures~\ref{fig:wallshearstress}($b$) and ($c$). The bottom panel in figure~\ref{fig:wallshearstress} shows the temporal evolution of the bulk polymer stretching $tr(\alpha)_b$ normalized by the maximum extensibility of polymers $b$ for various polymer concentrations. Interestingly, the bulk polymer stretching of all polymer concentrations is almost identical and increases very slowly until a sharp increase begins at almost the same time when the wall shear stress also starts to increase sharply.

\begin{figure}
\begin{center}\includegraphics[width = 5.2in]{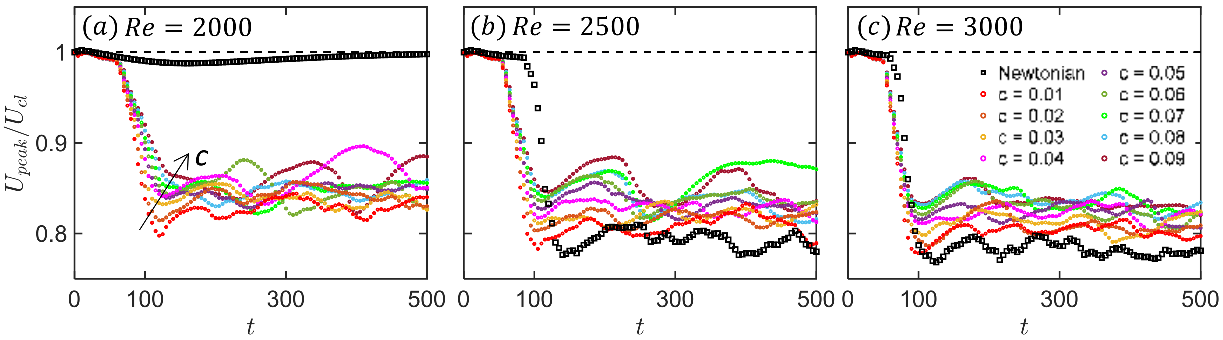}
\caption{Temporal evolution of the peak velocity normalized by the laminar centerline velocity for the perturbation amplitude $A=0.07$ at (a) $Re=2000$, (b) $Re=2500$, and (c) $Re=3000$: laminar state (black dashed line), Newtonian flow (black squares), and viscoelastic flow of various polymer concentrations (coloured symbols).
\label{fig:peakvel}}
\end{center}
\end{figure}

As an alternative to characterize the transition to turbulence, the distortion of the mean velocity profile has also been utilized for Newtonian flows, as its relationship with the formation of vortical structures during transition has been well established \citep{Lemoult2012}. Figure~\ref{fig:velprofs2} shows snapshots of the mean velocity profile $U_m$ normalized by the Newtonian laminar centerline velocity $U_{cl}$ at $Re = 2500$ at four different time instants for Newtonian and viscoelastic ($c = 0.03$) flows along with the base laminar profile as a reference. At $t = 55$, the velocity profile of both flows appears to be close to the base laminar profile. At the peak of the strong burst for the viscoelastic flow ($t = 90$), the deformation of the mean velocity profile is evident, while its Newtonian counterpart remains almost unchanged up to $t = 100$. At the peak of the strong burst for the Newtonian flow ($t = 125$), however, a severe deviation from the laminar profile is observed for both flows. Once the fully-turbulent state is reached ($t = 300$), the viscoelastic profile is closer to the laminar profile, suggesting drag reduction by polymers. For a further investigation, figure~\ref{fig:peakvel} shows the temporal evolution of the peak velocity $U_{peak}$ normalized by the centerline velocity $U_{cl}$ for Newtonian and viscoelastic flows of various polymer concentrations at $Re = 2000$, $2500$, and $3000$ perturbed with $A=0.07$. The base laminar state is also included for which $U_{peak}/U_{cl}=1$. Similarly, the response of the flow can be equally distinguished by cases (a) and (b) when utilizing $U_{peak}/U_{cl}$. For $Re=2000$ (fig.~\ref{fig:peakvel}$a$), the Newtonian flow departs slightly from the laminar state; however, the transition is not achieved and the flow remains laminar (case a), whereas the viscoelastic flows of all polymer concentrations deviate from the laminar state and continue the transitional path to turbulence (case b). Once the transition to turbulence is established, the velocity ratio quickly decreases until a fully-turbulent state is reached. For $Re\geq2500$, both Newtonian and viscoelastic flows respond following the path of case (b). The earlier departure from the laminar state ratio of 1 is clearly observed for viscoelastic flows in figure~\ref{fig:peakvel}($b$), as in the wall shear stress (fig.~\ref{fig:wallshearstress}$b$). The similarity in the start of transition in Newtonian and viscoelastic flows at $Re=3000$ can also be confirmed by figure~\ref{fig:peakvel}($c$).

\begin{figure}
\begin{center}\includegraphics[width = 3.9in]{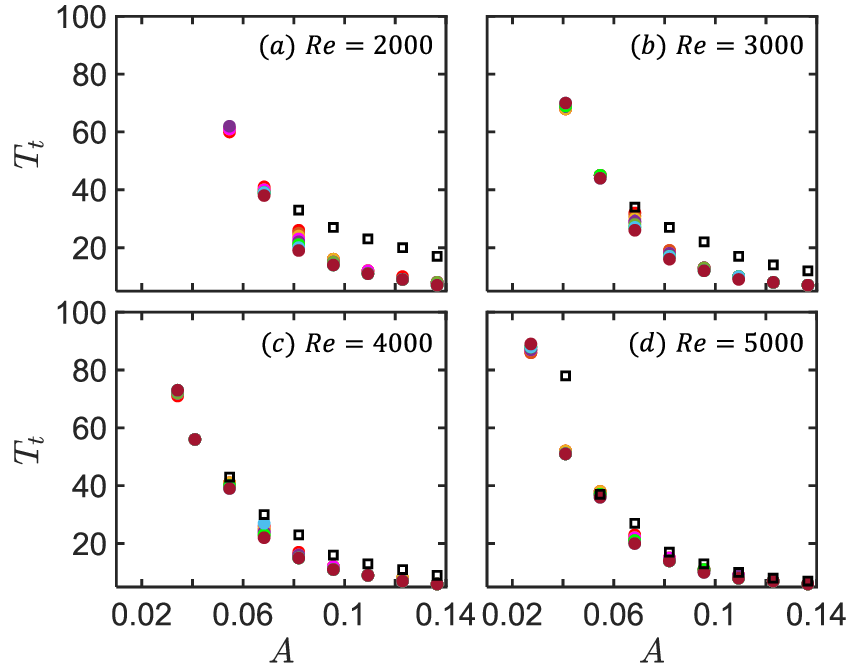}
\caption{Transition time as a function of perturbation amplitude $A$ at ($a$) $Re=2000$, ($b$) $Re=3000$, ($c$) $Re=4000$, and ($d$) $Re=5000$: Newtonian flow (black open squares) and viscoelastic flow of various polymer concentrations (coloured solid circles). Refer to the legend in figure \ref{fig:wallshearstress} for colours for polymer concentration.
\label{fig:trantimepert}}
\end{center}
\end{figure}

\subsection{Transition time: onset of transition}\label{subsec:trantime}
Now, we proceed to investigate the time for the onset of transition. The transition time $T_t$ is defined as the time at which the wall shear stress reaches 105\% of the base laminar value ($\tau_w = 2.1$). The sensitivity to the chosen threshold value was examined by utilizing different threshold values, such as 110\% and 115\%, showing almost identical trends with an upward shift in the time it takes for each case to reach the threshold criteria. In addition, the sensitivity to the chosen parameter was also examined by utilizing the peak velocity, such as figure~\ref{fig:peakvel}, with the threshold of 95\% of the base laminar value, showing an almost identical trend. It should be noted that the aforementioned criteria for the onset of transition can only be detected in cases where the perturbation amplitude is strong enough to trigger transition for both Newtonian and viscoelastic flows.

\begin{figure}
\begin{center}\includegraphics[width = 5.3in]{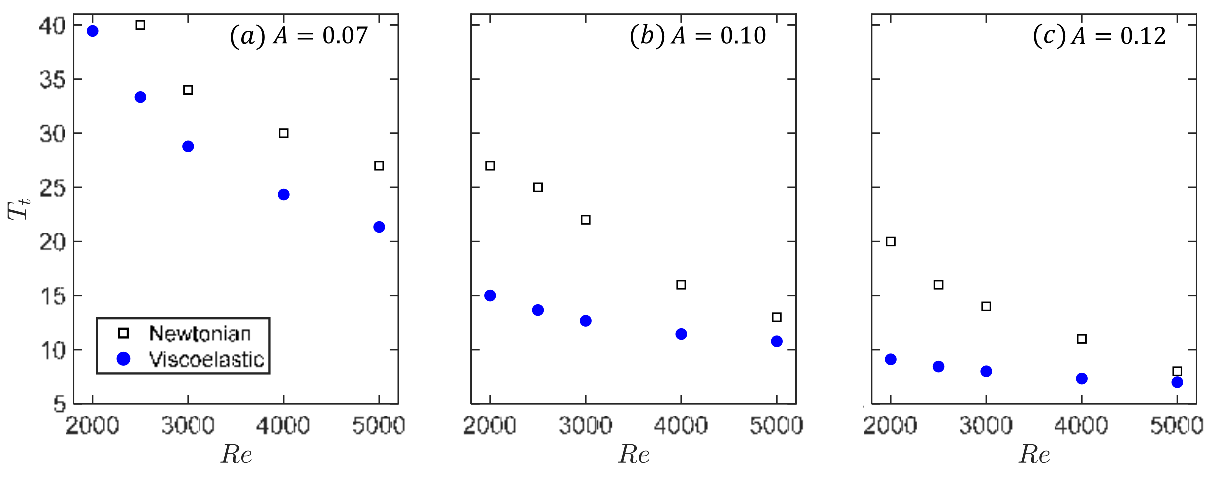}
\caption{Transition time as a function of Reynolds number $Re$ for the perturbation amplitude ($a$) $A = 0.07$, ($b$) $A = 0.10$, and ($c$) $A = 0.12$ for Newtonian flow (black open squares) and viscoelastic flow ($c=0.03$; blue solid circles).
\label{fig:trantimeRe}}
\end{center}
\end{figure}

\begin{figure}
\begin{center}\includegraphics[width = 3.5in]{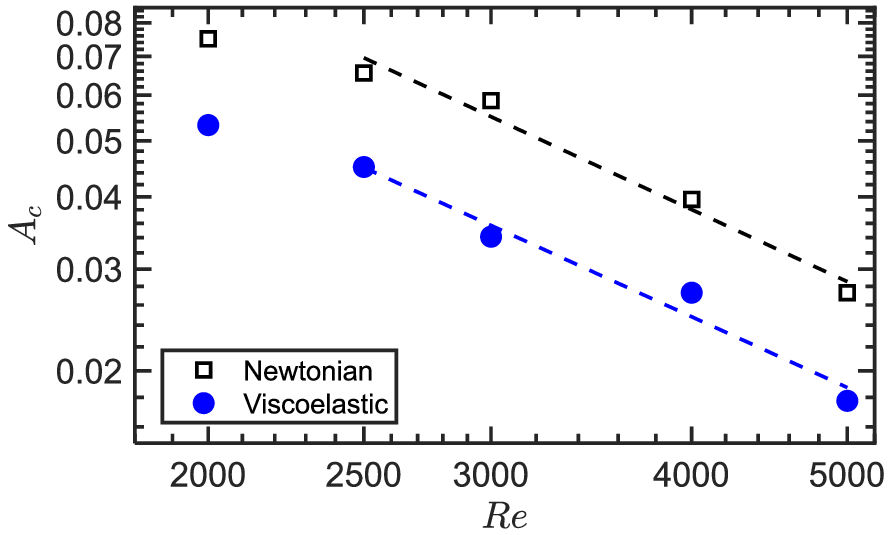} 
\caption{Log-log plot of the stability curve for the onset of transition to turbulence in the range of $2000\le Re\le5000$. Newtonian flow (black open squares) and viscoelastic flow ($c = 0.03$) (blue solid circles). For $Re \ge 2500$, the minimum or critical perturbation amplitude $A_c$ follows a power-law scaling of $A_c = O(Re^{\gamma})$, where $\gamma \approx -1.25$ for both Newtonian and viscoelastic flows.
\label{fig:stabcurve}}
\end{center}
\end{figure}

Figure~\ref{fig:trantimepert} shows the transition time for Newtonian and viscoelastic flows of various polymer concentrations at Reynolds number up to $5000$ as a function of perturbation amplitude. As expected, the overall trend appears to be similar for both Newtonian and viscoelastic flows, where the transition time decreases with increasing perturbation amplitude and Reynolds number. The earlier transition for viscoelastic flows is also confirmed by a smaller $T_t$ regardless of polymer concentration (i.e. transition is initiated at an earlier time than Newtonian flow). Interestingly, it seems that the polymer concentration has a negligible effect on the transition time for all $Re$. Yet, as clearly observed in all Reynolds numbers studied, the main difference is that the viscoelastic flow requires a lower perturbation amplitude to trigger the transition in comparison to its Newtonian counterpart. For example, figure~\ref{fig:trantimepert}($a$) shows that at $Re = 2000$, the lowest or critical perturbation amplitudes to trigger transition are $A \approx 0.08$ and $A \approx 0.05$ for the Newtonian and viscoelastic flows, respectively. Increasing the Reynolds number lowers the critical perturbation amplitude for both Newtonian and viscoelastic flows. In addition, the difference between the Newtonian and viscoelastic transition times $\Delta T_t$ gets smaller with increasing $Re$. In order to better explore the effect of Reynolds number, we replot the transition time as a function of Reynolds number at a given perturbation amplitude, as shown in figure~\ref{fig:trantimeRe}. For viscoelastic flows, the polymer concentration $c = 0.03$ is only used as the polymer concentration appears to have no impact on the transition time. At three perturbation amplitudes considered, the transition time decreases monotonically for both flows with increasing Reynolds number. For a relatively weak perturbation amplitude $A = 0.07$ (fig.~\ref{fig:trantimeRe}$a$), $\Delta T_t \approx 6$ and this value remains almost constant with increasing Reynolds number. However, for relatively strong perturbation amplitudes of $A > 0.07$ (figs.~\ref{fig:trantimeRe}$b,c$), $\Delta T_t$ becomes more significant for lower Reynolds numbers but gets much smaller with increasing Reynolds number. For $A = 0.12$ (fig.~\ref{fig:trantimeRe}$c$), the transition time of the viscoelastic flow barely decreases with Reynolds number, and $\Delta T_t$ is almost negligible at $Re = 5000$. It is worth noting that at the given Reynolds number, the transition time decreases with the perturbation amplitude. However, the level of drag reduction achieved by the same polymer parameters during sustained turbulence would be very similar and insensitive to the transition time.

\subsection{Stability curve: critical perturbation amplitude}\label{subsec:stability}
Next, we present the finite-amplitude thresholds for the perturbation to start triggering the transition or the critical perturbation amplitude $A_c$ below which no transition occurs. Figure~\ref{fig:stabcurve} shows $A_c$ for Newtonian and viscoelastic ($c = 0.03$) flows on a log-log scale. As expected from the previous studies for Newtonian flows (e.g., \cite{Hof2003,Lemoult2012}), our Newtonian flow (black open squares) clearly follows a power-law scaling of $Re^\gamma$ for $Re \ge 2500$. As was observed by \cite{Lemoult2012}, it is also observed that the asymptotic regime is not reached at lower Reynolds numbers for $Re < 2500$. The critical exponent $\gamma$ for the Newtonian flow is very close to $-1.25$, which has also been reported for transition in plane Poiseuille flows triggered by a perturbation leading to streamwise vortices \citep{Chapman2002}. Most interestingly, the viscoelastic flow also follows almost the same power-law scaling as the Newtonian flow with $\gamma \approx -1.25$ for $Re \ge 2500$. It can also be observed that the viscoelastic $A_c$ is smaller than the Newtonian $A_c$, which suggests that smaller finite perturbation amplitudes are sufficient to trigger transition for viscoelastic flows compared to its Newtonian counterparts. It should be noted that the same scaling law with a constant exponent is found for different perturbation fields generated by \textit{ChannelFlow}, while the asymptotic lines are shifted upward or downward depending on the characteristics of the perturbation structure (see appendix).

\begin{figure}
\begin{center}\includegraphics[width = 4.7in]{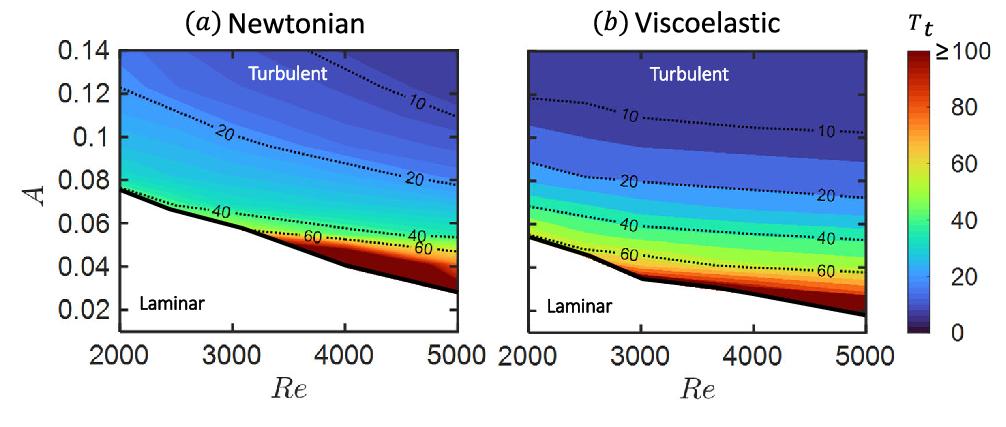}
\caption{Colour contour of the transition time $T_t$ due to a finite amplitude of the perturbation $A$ at Reynolds number $Re$ for ($a$) Newtonian flow and ($b$) viscoelastic ($c = 0.03$) flow. The thick black line separates the basins of attraction of laminar and turbulent flows.
\label{fig:contour}}
\end{center}
\end{figure}

For a comprehensive understanding of transition to turbulence and a direct comparison between Newtonian and viscoelastic flows, figures~\ref{fig:contour}($a$) and ($b$) present the state diagram of the transition time $T_t$ in a space of the perturbation amplitude $A$ and Reynolds number $Re$ for Newtonian and viscoelastic ($c = 0.03$) flows, respectively. As can be seen, there are two distinct regions, laminar and turbulence, separated by the laminar-turbulent boundary (thick black line) separating the basins of attraction of laminar and turbulent flows \citep{Schneider2007,Duguet2008}. This boundary indeed represents the critical perturbation amplitude $A_c$ in figure~\ref{fig:stabcurve}. Since there are theoretical arguments that the so-called exact coherent states (ECS) form a part of the basin boundary \citep{Kawahara2005, Wang2007}, the dynamics on or near this boundary could play an important role in finding new ECSs in viscoelastic flows or EIT \citep{Page2020, Dubief2022, Buza2022}, which will be a subject of interesting future work.

Clearly, this boundary is shifted down for the viscoelastic flow, suggesting that the laminar region is shrunk by polymers. It also suggests the reduction in the perturbation magnitude required to trigger transition for viscoelastic flows. A shorter transition time of the viscoelastic flow can be readily identified in comparison to the Newtonian flow at the given Reynolds number and perturbation amplitude. Interestingly, the decay rate of the transition time with respect to the Reynolds number shows distinct trends. As can be seen in lines of the constant $T_t$ values from 60 to 10, the slope of these lines gets steeper for the Newtonian flow, while the lines become level off for the viscoelastic flow. It could suggest that at very high perturbation amplitudes, the effect of polymers on the transition time or the early stage of transition is almost the same, regardless of Reynolds number, which can also be confirmed by figure \ref{fig:trantimeRe}($c$).

\begin{figure}
\begin{center}\includegraphics[width = 5in]{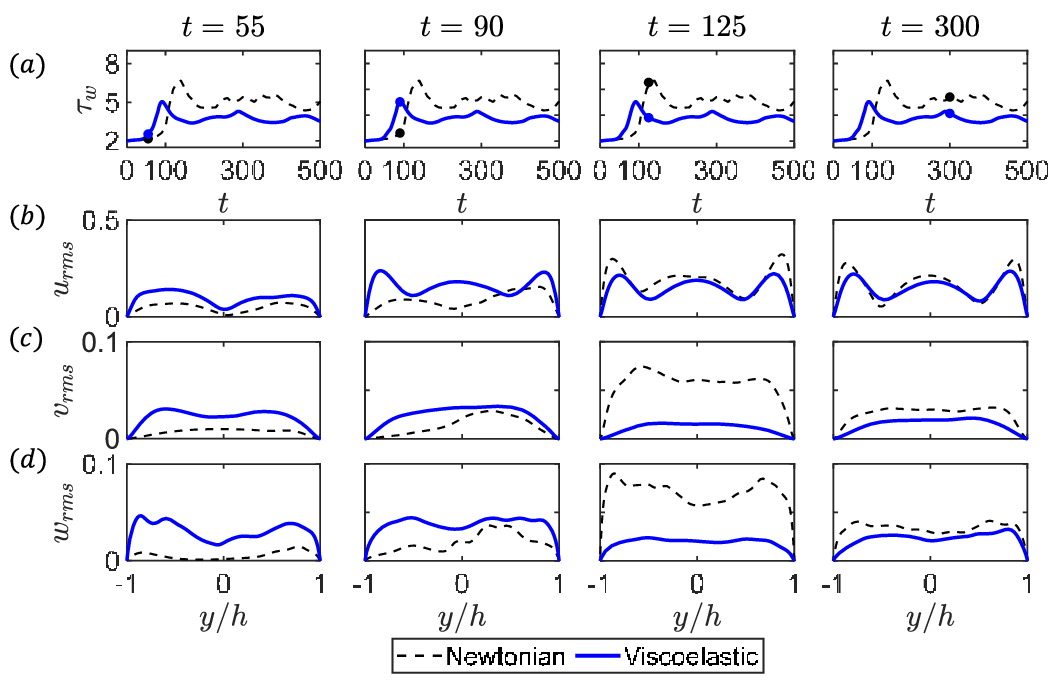}
\caption{($a$) Temporal evolution of the wall shear stress $\tau_w$ at $Re = 2500$ for the perturbation amplitude $A = 0.07$ for Newtonian flow (black dashed line) and viscoelastic flow ($c = 0.03$; blue solid line) with the time instants $t = 55, 90, 125,$ and $300$. Snapshots of the ($b$) streamwise, ($c$) wall-normal, and ($d$) spanwise velocity fluctuations at each time instant shown in ($a$). Also see the accompanying supplementary movie.
\label{fig:snapsumm}}
\end{center}
\end{figure}

\subsection{Mechanism: velocity fluctuations, flow structures, and polymer dynamics}\label{subsec:velflucts}
We now attempt to elucidate the mechanisms behind the earlier transition triggered by polymers in viscoelastic flows. Figure~\ref{fig:snapsumm}($a$) shows the evolution of the wall shear stress at $Re = 2500$ for Newtonian and viscoelastic ($c = 0.03$) flows at four different time instants. Figures~\ref{fig:snapsumm}($b$-$d$) show snapshots of the velocity fluctuations in the streamwise $u_{rms}$, wall-normal $v_{rms}$, and spanwise $w_{rms}$ directions at each time instant for both flows. An observation to draw from figures~\ref{fig:snapsumm}($c$,$d$) at $t = 55$ is that changes in the wall-normal velocity fluctuation $v_{rms}$ and spanwise velocity fluctuation $w_{rms}$ in the viscoelastic flow, with respect to its Newtonian counterparts, are somewhat more significant compared to the streamwise velocity fluctuation $u_{rms}$. For the Newtonian case, $v_{rms}$ is significantly lower near the wall and $w_{rms}$ is barely observed compared to ones for the viscoelastic flow. This suggests that the wall-normal and spanwise directions are destabilized earlier by polymers, eventually promoting an earlier transition. In other words, an early transition observed in the viscoelastic flow is attributed to the early destabilization of the wall-normal and spanwise directions due to polymers. As the transition process proceeds, $v_{rms}$ and $w_{rms}$ of the Newtonian flow take over and continue to be larger than ones of the viscoelastic flow. Once the transitional period of both flows has passed ($t =300$), a fully-sustained turbulent regime begins, where the expected characteristics of the drag-reduced flow are observed that $v_{rms}$ and $w_{rms}$ of the viscoelastic flow are lower than ones of the Newtonian flow \citep{Li2006}. This whole process can be seen in the accompanying supplementary movie.

\begin{figure}
\begin{center}\includegraphics[width = 4.8in]{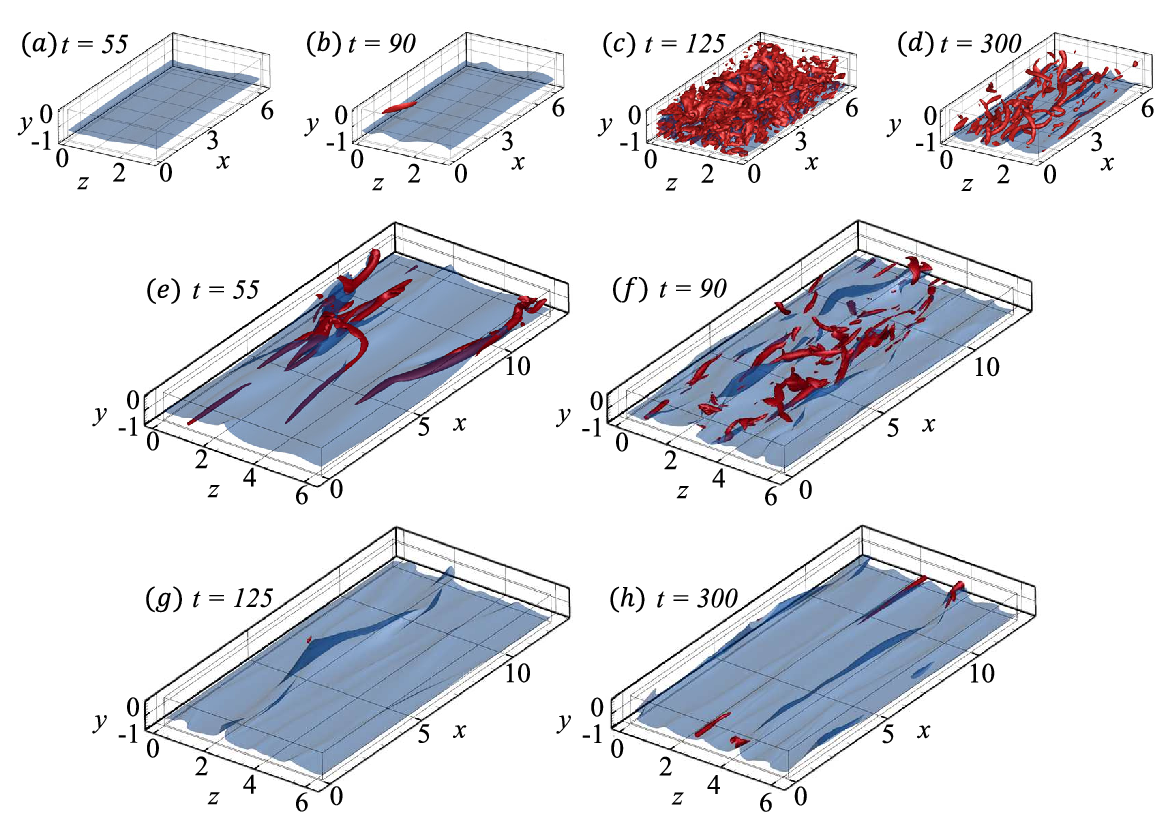}
\caption{Temporal evolution of vortical structures for the bottom half-channel: ($a-d$) for Newtonian and ($e-h$) viscoelastic ($c=0.03$) flows at $Re = 2500$ triggered by the perturbation amplitude $A = 0.07$. The red tubes are isosurfaces at constant vortex strength $Q = 0.216$, and the blue contours are isosurfaces of constant streamwise velocity. The maximum strengths are $(a)$ 0.018, $(b)$ 1.29, $(c)$ 14.64, $(d)$ 4.25, $(e)$ 4.49, $(f)$ 2.39, $(g)$ 0.25 and $(h)$ 0.63. For reference, the maximum vortex strength for sustained turbulence are $Q = 4.42$ and $Q = 0.48$ for Newtonian and viscoelastic flows, respectively. Also see the accompanying supplementary movies.
\label{fig:vortex}}
\end{center}
\end{figure}

This destabilization mechanism is further studied by estimating the effect of viscoelasticity on flow structures. Vortex identification is performed by utilizing the so-called $Q$-criterion \citep{JeongHussain1995} for which $Q=\frac{1}{2} (||\boldsymbol{\Omega}||^2-||\textbf{\textit{S}}||^2)$ is computed, where $\textbf{\textit{S}}\equiv\frac{1}{2}(\nabla\textbf{\textit{u}}+\nabla\textbf{\textit{u}}^T)$ is the rate of strain tensor and $\boldsymbol{\Omega}\equiv\frac{1}{2}(\nabla\textbf{\textit{u}}-\nabla\textbf{\textit{u}}^T)$ is the vorticity tensor. Regions of $Q > 0$ correspond to the areas of strong vortical motions. Figure~\ref{fig:vortex} shows snapshots of these vortical structures for the lower half of the channel for Newtonian flow ($a-d$) and viscoelastic flows of $c = 0.03$ ($e-h$) at different time instants that can also be seen in the evolution of wall shear stress in figure~\ref{fig:snapsumm}($a$). The red tubes represent isosurfaces of $Q = 0.216$, which is approximately half of the maximum of the sustained viscoelastic turbulent flow for this particular case. For comparison, the half of the maximum of the sustained Newtonian turbulent flow is $Q = 2.21$. The blue contours correspond to isosurfaces of constant streamwise velocity, which could represent a spatial structure for the streak.

Figures~\ref{fig:vortex}($a$) and ($e$) show the flow structures at the beginning stage of the transition process ($t = 55$) for Newtonian and viscoelastic flows, respectively. As polymers destabilize the flow earlier and result in the rapid energy growth, the streamwise-elongated red tubes or quasi-streamwise vortices start to form around wavy, uplifted blue isosurfaces, whereas its Newtonian counterpart remains undisturbed. This wavy, uplifted structure indeed signifies the form of low-speed streaks, which is one of the key ingredients for the self-sustaining process \citep{Waleffe1997}. It should be noted that given the flow regime of the current study, the spanwise-oriented structure that are dominantly observed in the EIT regime is unlikely to be observed. As the viscoelastic flow reaches its bursting peak at $t = 90$, figure \ref{fig:vortex}($f$) shows a relatively larger population of vortex cores formed around more enhanced low-speed streaks. After the bursting peak, the vortices are then quickly dampened as polymers work against them, entering a drag-reduced turbulent regime (fig.~\ref{fig:vortex}$h$ at $t = 300$). In comparison, when its Newtonian counterpart reaches its bursting peak at $t = 125$ (fig.~\ref{fig:vortex}$c$), more heavily populated vortex cores are formed across almost the entire domain, where the characteristics of the vortical structures are hard to identify. After the strong bursting peak, however, some of the quasi-streamwise vortices are observed as seen in figure~\ref{fig:vortex}($d$).

In addition to the velocity fluctuations and flow structures, polymer dynamics can also provide the underlying mechanism behind the early transition. As the more stretched polymer state ties into the more unstable flow state \citep{XiGraham2010, Graham2014}, we can refer to the bulk polymer stretching in figure~\ref{fig:wallshearstress}. In the early transition stage, the polymers start to stretch slowly as they interact with the flow. As they keep stretching, polymers continue to destabilize the flow to enhance the velocity fluctuations. The transition eventually occurs when the bulk polymer stretching reaches a certain value, depending on the Reynolds number and perturbation amplitude, after which the bulk polymer stretching starts to increase sharply, as seen in figure~\ref{fig:wallshearstress}.

In short, the destabilization mechanism of polymers during transition is attributed to the early amplification of the wall-normal and spanwise velocity fluctuations and the formation of the quasi-streamwise vortices around low-speed streak, all of which facilitate an early transition. Interestingly, the vortical structures observed in both Newtonian and viscoelastic flows also support the power exponent $\gamma = -1.25$ in the power-law scaling of the critical perturbation amplitude in figure~\ref{fig:stabcurve}, as also observed in the previous study \citep{Chapman2002}. It should be noted, however, that a perturbation leading to different flow structures could provide a power-law scaling with a different exponent. In addition, this study focuses on a dilute polymer solution ($c < 0.1$) and low elasticity ($El < 0.02$), where an early transition is only observed. Thus, it could suggest a subject of future work for the transition to turbulence in polymer solutions at semi-concentrated or highly concentrated regimes above the overlap concentration and at high elasticity ($El \gg 0.02$) such as within the EIT regime, where different transition scenarios or flow structures could arise \citep{Datta2022}.

\section{Conclusion}\label{sec:conclusion}
Direct numerical simulations of dilute polymer solutions with a FENE-P model were performed to investigate the effect of polymers on the laminar-turbulent transition in plane Poiseuille (channel) flow. Starting from a base laminar state, which is disturbed with a small finite-amplitude perturbation, the short stable period was observed for both Newtonian and viscoelastic flows at the beginning followed by the linear and nonlinear evolution of the disturbance energy. However, the viscoelastic flow experiences a shorter duration of the stable period, hinting at a destabilizing effect of the polymers during the early stages of transition. Also, we observed that the viscoelastic flow requires a smaller amplitude of perturbation to trigger transition, whereas the Newtonian counterpart remains undisturbed until a larger amplitude is utilized to trigger transition. We show that the transition time $T_t$, defined as the onset time of transition, decreases with increasing perturbation amplitude. As polymers promote an early transition, the Newtonian flow exhibits a higher $T_t$ than the viscoelastic flow, but this difference becomes less pronounced as $Re$ is increased. Interestingly, the polymer concentration studied ($0.01 \le c \le 0.09$) barely has an effect on the transition time. However, the higher the polymer concentration, the lower the magnitude of the strong burst, suggesting that higher polymer concentrations exhibit enhanced drag-reducing behavior after the onset of transition.

We then investigated the critical perturbation amplitude $A_c$, which is the minimum amplitude to trigger the transition. The viscoelastic flow shows a smaller $A_c$ than its Newtonian counterpart, suggesting that polymers kick in the destabilizing effect early. Similar to previous studies for Newtonian transition, the critical amplitude of our Newtonian flow follows a power-law scaling of $Re^\gamma$ for $Re \ge 2500$, where $\gamma \approx -1.25$. More interestingly, a viscoelastic flow also follows almost the identical power-law scaling as the Newtonian flow. Both flows display almost the same critical exponent of $\gamma\approx-1.25$ for $Re \ge 2500$. This critical exponent implies that the perturbation of the current study leads to the formation of quasi-streamwise vortices \citep{Chapman2002}.

The early transition or destabilization effect triggered by polymers is further investigated. During the stable period at the beginning stage of transition, the wall-normal and spanwise velocity fluctuations start to grow early in the viscoelastic flow, compared to those of the Newtonian flow. Hence, the polymers appear to destabilize these components of the velocity fluctuation quickly, promoting an earlier transition. Once the fully-turbulent stage is reached after the strong burst, however, the drag-reducing behaviour of polymers is observed, where the wall-normal and spanwise velocity fluctuations for the viscoelastic flow are reduced and become smaller than those of the Newtonian flow. This destabilizing effect of polymers was also confirmed by considering the bulk polymer stretching and visualizing the flow structure, where the vortical motions are shown up earlier around low-speed streaks for a viscoelastic flow. Interestingly, it is possible to see the formation of quasi-streamwise vortices for both flows, which supports the critical exponent of $\gamma\approx-1.25$ for the power-law scaling of the critical amplitude.

This study further complements the previous studies on the laminar-turbulent transition by providing the power-law scaling of the critical perturbation amplitude for both Newtonian and viscoelastic flows, which has been mostly unexplored. The destabilizing effect of polymers during the early stage of transition is consistent with the effects of a low elasticity number and polymer concentration, as is the case of our study. Moving forward, the robustness of power-law scaling on different perturbation structures should be further investigated. Additionally, higher elasticity numbers and polymer concentrations may lead to different transition dynamics and mechanisms, which will be a subject of interesting future work.

\section*{Acknowledgements}
Research is supported by the U.S. Department of Energy, Office of Science, Office of Basic Energy Sciences under award number DE-SC0022280 (modeling studies) and by the National Science Foundation under award numbers OIA-1832976 and CBET-2142916 (CAREER) (computational studies). The direct numerical simulation code used was developed and distributed by John Gibson at the University of New Hampshire. The authors also acknowledge the computing facilities used at the Holland Computing Center at the University of Nebraska-Lincoln.
\\
\\
Declaration of interests: The authors report no conflict of interest.

\section*{Appendix. Different perturbation forms}
\begin{figure}
\begin{center}\includegraphics[width = 5.2in]{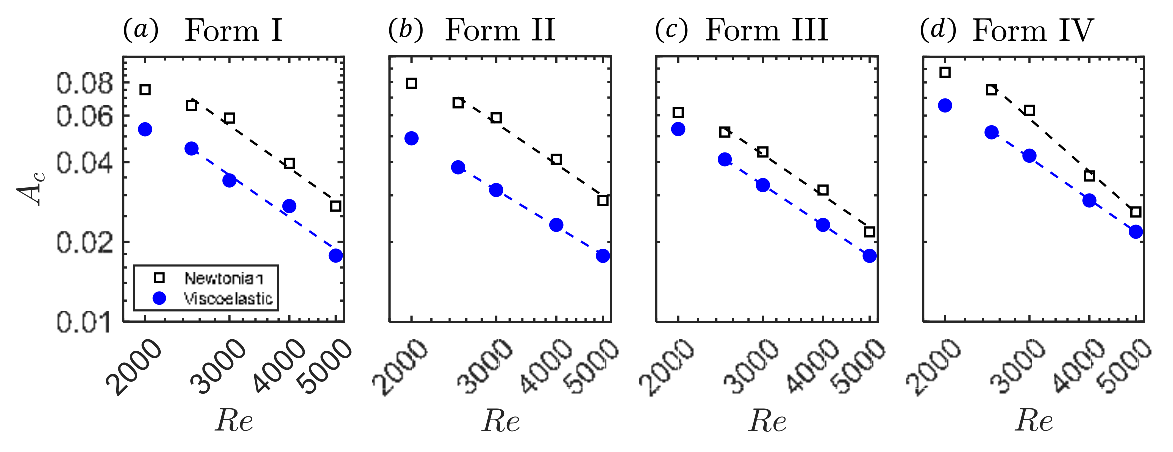}
\caption{Log-log plot of the stability curve for the onset of transition to turbulence in the range of $2000 \leq Re \leq 5000$ of the different perturbations ($a$) Form I (current study), ($b$) Form II, ($c$) Form III, and ($d$) Form IV. Each form is meant to display different vortical structures in terms of the strength and shape. Newtonian flow (black open squares); viscoelastic flow ($c = 0.03$) (blue solid circles). For $Re \geq 2500$, the minimum or critical perturbation amplitude $A_c$ follows a power-law scaling of $A_c = O(Re^\gamma)$, where $\gamma \approx -1.25$ for all the cases, except for a Newtonian case in Form IV of $\gamma \approx -1.50$.
\label{fig:diff_seeds}}
\end{center}
\end{figure}

Figure~\ref{fig:diff_seeds} shows the critical perturbation amplitude $A_c$ for Newtonian and viscoelastic ($c = 0.03$) flows on a log-log scale for various perturbation forms, where Form I is the perturbation used in the current study. These perturbations were generated by $ChannelFlow$, which allows us to create different forms of the perturbation field in terms of different vortical structures and strengths. It is observed that the same power-law scaling of $A_c = O(Re^{\gamma})$ with $\gamma\approx -1.25$ is found for both Newtonian and viscoelastic flows with all perturbation forms studied for $Re \geq 2500$, while asymptotic lines are shifted upward or downward depending on different perturbation forms. It should be noted, however, that a Newtonian flow of Form IV exhibits a little higher exponent of $\gamma \approx -1.50$. Although the different forms of the perturbation leads to the same stability scaling, a detailed analysis of the transition dynamics due to each different perturbation form is necessary but beyond the scope of the current study, which will be a subject of future work.

\bibliographystyle{jfm}
\bibliography{refs}

\end{document}